\documentclass[11pt]{article}
\usepackage{hyperref}
\usepackage{graphicx}
\pdfoutput=1
\newcommand{\fref}[1]{figure~\ref{#1}}

\newcommand{\px}{\ensuremath{\,\mbox{pixel}}}

\begin{document}
\title{PIV Analysis of Ludwig Prandtl's Historic Flow Visualization Films}
\author{C. Willert$^1$, J. Kompenhans$^2$ \\[9pt]
$^1$) Institute of Propulsion Technology,\\
    German Aerospace Center (DLR),
    51170 K\"oln, Germany\\[6pt]
$^2$) Institute of Aerodynamics and Flow Technology,\\
    German Aerospace Center (DLR),
    37073 G\"ottingen, Germany}
\maketitle
%

\begin{abstract}

Around 1930 Ludwig Prandtl and his colleagues O. Tietjens and W. Müller published two films with visualizations of flows around surface piercing obstacles to illustrate the unsteady process of flow separation. These visualizations were achieved by recording the motion of fine particles sprinkled onto the water surface in water channels. The resulting images meet the relevant criteria of properly seeded recordings for particle image velocimetry (PIV). Processing these image data with modern PIV algorithms allows the visualization of flow quantities (e.g. vorticity) that were unavailable prior to the development of the PIV technique. The accompanying fluid dynamics video consists of selected original film sequences overlaid with visualizations obtained through PIV processing.

\end{abstract}


\section{Introduction}

Probably some of the oldest time-resolved particle image velocimetry (PIV) image sequences were acquired by Prandtl and his colleagues Tietjens and M\"{u}ller \cite{IWF_C1:1936,IWF_C2:1936}.
By the late 1920's Prandtl and his colleagues began recording visualizations of unsteady free surface flows created by facilities such as water flumes using cinematic camera equipment. The visualization was achieved by means of small particles (aluminum powder, ferrous mica ("Eisenglimmer"), or lycopodium powder) scattered on the surface following a method developed earlier by Prof. F.~Ahlborn in Hamburg [citation needed]. The films were mainly intended for instructional purposes to illustrate the process of flow separation. The recordings on 16~mm film were acquired at 20~frames per second either in a laboratory-stationary frame or moving at the same speed as a surface-piercing object like an airfoil or cylinder. Contrary to the frequently used time integrated single recordings which resulted in particle streak images (see e.g. \cite{Prandtl:1927}), the time resolved recordings allowed, for instance, the clear observation of upstream flow inside separation bubbles, a subject of considerable dispute at the time as manifested by a lengthy discussion between Prandtl, Ahlborn and other scientists following a presentation by Ahlborn \cite{Ahlborn:1927} at the 1927 congress of the Wissenschaftliche Gesellschaft f\"{u}r Luftfahrt (WGL) in Wiesbaden, Germany.

\begin{figure}
\includegraphics*[width=.48\textwidth]{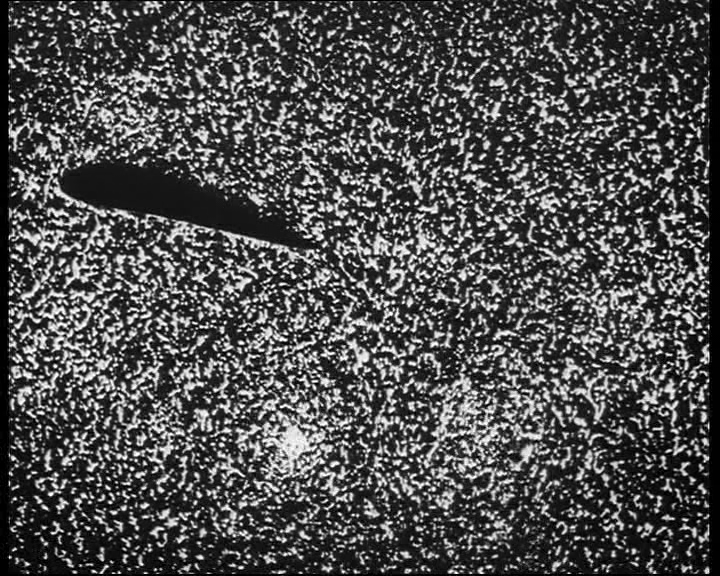}
\includegraphics*[width=.48\textwidth]{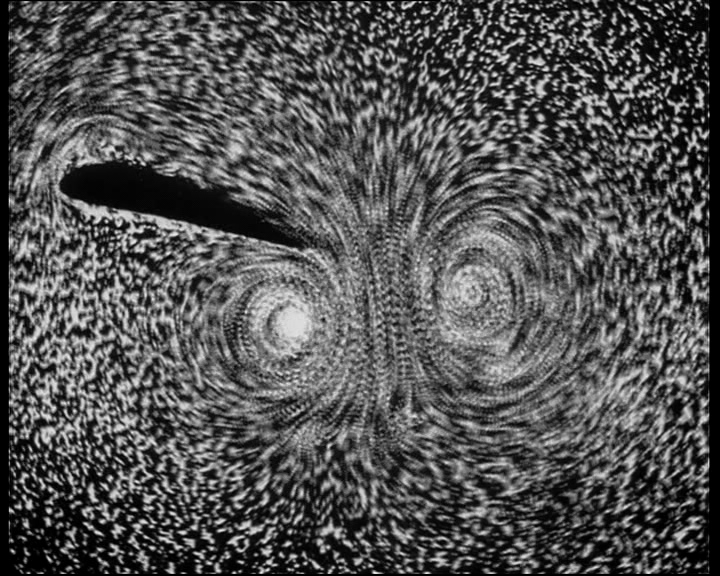}
\caption[]{Single movie frame (left) and composite of five frames (right) of an airfoil that is impulsively translated from right to left (from \protect\cite{IWF_C1:1936})\label{fig:prandtl_wing}}
\end{figure}

The original visualization films of Prandtl were recently made available in the form of a Digital Video Disk (DVD) \cite{IWF_DVD:2009} by the "Institut für wissenschaftlichen Film" (IWF, Institute for Scientific Film) in G\"{o}ttingen. An exemplary frame of the acquired movies are provided in \fref{fig:prandtl_wing} for a airfoil that is translated by about one chord length. Remarkable about these image sequences is the very homogenous particle distribution and the practical absence of particle image streaking, that makes it difficult to discern the nature of the flow from a single image. In effect the images exhibit the recommended particle image density and distribution for PIV.

\begin{figure}
\includegraphics*[width=.5\textwidth]{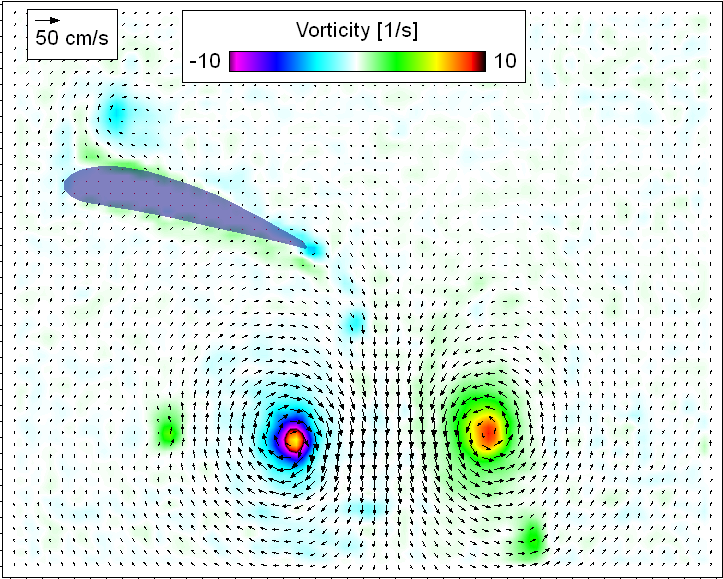}
\caption[]{Recovered velocity and vorticity field for image shown in \protect\fref{fig:prandtl_wing}\label{fig:prandtl_wing_vort}}
\end{figure}

Given the time series of surface flow visualizations it is a simple and elucidating exercise to use today's computer based PIV algorithms to extract the displacement and corresponding velocity and vorticity fields. An exemplary result is provided in \fref{fig:prandtl_wing_vort} for the image shown in \fref{fig:prandtl_wing}. While the original images were recorded on 16~mm film, the digitized version has a spatial resolution of only $720 \times 576 \px$ (PAL video format). The images can be reliably analyzed with image samples as small as $16 \times 16 \px$ yielding $45 \times 36$ discrete vectors. Smaller sample sizes introduce artifacts due to both the finite seeding density (insufficient particle images per sample) as well as due  to compression artifacts of DVD encoding. Contrast enhancing measures prior to PIV processing were not necessary. The interrogation algorithm is based on a three level pyramid approach starting with samples of $64 \times 64 \px$ and refining to $16 \times 16 \px$ \cite{PIVBook2:2007}. Full image deformation using B-spline interpolation (3rd order) is applied after each validated interrogation pass. Validation using a normalized median test \cite{Westerweel2005} and range filtering indicates a validation rate greater than 95\%. Detected outlier vectors are replaced using bilinear interpolation.

The recovered PIV displacement data can only roughly be converted to velocity data since the actual scaling cannot be extracted from the footage provided on the DVD. A few sequences briefly show the hand of an operator adjusting the flow facility which allows a rough approximation of the length scale. The 'pulse delay' is directly obtained by the fixed frame rate of 20~Hz which corresponds to 50~milliseconds between recordings. From this the flow velocities in the original experiments are estimated to have been in the range $10 - 200$ cm/s.

Further analysis of the displacement (velocity) data show a periodic jitter which is caused by frame-to-frame image registration jitter, most likely introduced during the original recording process (or possibly also during the later digitization). The effect is most noticeable in region of quiescent flow where displacement vectors tend to 'dance' unison. This is the main reason the accompanying fluid dynamics video does not present displacement vector plots. Since this displacement jitter is a global effect it can -- in principle -- be removed with further processing. It is also removed when finite difference are applied to the displacement fields and therefore does not appear in the presented vorticity maps.

\begin{figure}
\includegraphics*[width=.5\textwidth]{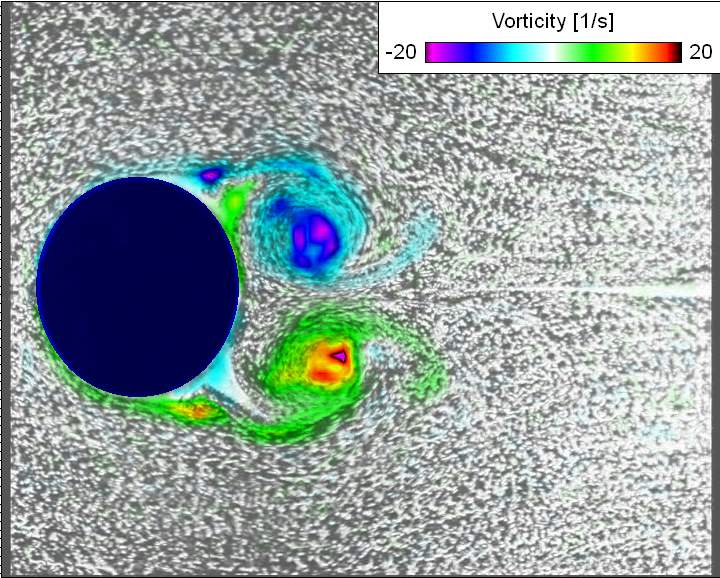}
\caption[]{Sample from the fluid dynamics video: original photographic recording overlaid with vorticity field obtained by differentiation of the PIV estimated flow field\label{fig:prandtl_cyl_vort}}
\end{figure}

Visualizations of the evolving vorticity fields comprise the main content of the fluid dynamics video and very nicely illustrate unsteady shedding, vortex formation and transition effects. A sample of a vorticity map overlaid on the original image data is presented for the starting flow downstream of a cylinder in \fref{fig:prandtl_cyl_vort}. In total four sequences are presented:

\begin{itemize}
  \item the starting flow downstream of a cylinder,
  \item the flow around an airfoil that is impulsively moved by about one chord length,
  \item the onset of flow separation on the suction side of a bluff body,
  \item the separated flow caused by a sharp-edged plate placed normal to the mean flow direction.
\end{itemize}

\bibliographystyle{cew_aps2010}
\bibliography{Biblio_CWillert}

\end{document}